# Single NV in nanodiamond for quantum sensing of protein dynamics in an ABEL trap

Iván Pérez[a], Anke Krueger[b,e], Jörg Wrachtrup[c,e], Fedor Jelezko[d,e], Michael Börsch*[a,e]

[a] Single-Molecule Microscopy Group, Universitätsklinikum Jena, Nonnenplan 2 - 4, 07743 Jena;
[b] Institut für Organische Chemie, Universität Stuttgart, Stuttgart;
[c] 3. Physikalisches Institut, Universität Stuttgart, Stuttgart;
[d] Institut für Quantenoptik, Universität Ulm, Ulm;
[e] Carl-Zeiss-Stiftung Zentrum für Quantenphotonik (QPhoton) Jena-Stuttgart-Ulm, Germany

## ABSTRACT

Enzymes are cellular protein machines using a variety of conformational changes to power fast biochemical catalysis. Our goal is to exploit the single-spin properties of the luminescent NV (nitrogen-vacancy) center in nanodiamonds to reveal the dynamics of an active enzyme complex at physiological conditions with the highest spatio-temporal resolution. Specifically attached to the membrane enzyme $F_oF_1$-ATP synthase, the NV sensor will report the adenosine triphosphate (ATP)-driven full rotation of $F_o$ motor subunits in ten consecutive 36° steps. Conformational dynamics are monitored using either a double electron-electron resonance scheme or $NV^-$ magnetometry with optical readout or using $NV^-$ relaxometry with a superparamagnetic nanoparticle as the second marker attached to the same enzyme. First, we show how all photophysical parameters like individual size, charge, brightness, spectral range of fluorescence and fluorescence lifetime can be determined for the $NV^-$ center in a single nanodiamond held in aqueous solution by a confocal anti-Brownian electrokinetic trap (ABEL trap). Stable photon count rates of individual nanodiamonds and the absence of blinking allow for observation times of single nanodiamonds in solution exceeding hundreds of seconds. For the proposed quantum sensing of nanometer-sized distance changes within an active enzyme, we show that local magnetic field fluctuations can be detected all-optically by analyzing fluorescence lifetime changes of the $NV^-$ center in each nanodiamond in solution.

## 1. INTRODUCTION

The large protein machine $F_oF_1$-ATP synthase provides living cells with the chemical "energy currency" adenosine triphosphate (ATP) [1, 2]. This membrane-integrated enzyme is found in all types of organisms from bacteria to fungi, from plants to animals and humans. The main enzymatic reaction is ATP formation from adenosine diphosphate (ADP) and phosphate, taking place sequentially in three synchronized catalytic binding sites on the $F_1$ domain of the enzyme. ATP synthesis is driven by an electrochemical potential difference of protons across a lipid bilayer membrane ("proton motive force" in most organisms) [3]. The proton motive force causes rotation of subunits in the membrane-embedded $F_o$ domain in a turbine-like movement that is transduced to the coupled rotary subunits γ and ε of the coupled $F_1$ domain. 120°-stepped rotation of the γ-subunit enforces the opening of one catalytic binding site for the release of newly-made ATP, enables catalysis of ATP formation within the second fully closed binding site, and supports the closing of the third binding site with bound ADP, bound phosphate and a $Mg^{2+}$ ion. The reverse chemical reaction ATP hydrolysis is strongly controlled by $F_oF_1$-ATP synthases. The hydrolysis of ATP to ADP and phosphate is associated with proton pumping across the membrane and is driven by subunit rotation in the opposite direction [4]. The sequential atomic changes in the three binding sites had been unraveled by x-ray crystallography in 1994 [5], and in 1997 γ-subunit rotation was demonstrated

...................................................................................................................................................

unequivocally by single-molecule videomicroscopy of a fluorescently labeled µm-long marker attached to the rotary γ-subunit [6]. Afterward, cryo-electron microscopy revealed a multitude of structural details of the entire $F_oF_1$-ATP synthase in different conformational states and from distinct biological origins [7-9]. The use of light-scattering gold nanoparticles as small reporters of subunit rotation enabled the angular-resolved detection of the dwell times of the individual catalytic steps (i.e., during ATP hydrolysis) with sub-millisecond time resolution [10-12].

We have taken an alternative approach to monitor subunit rotation of a single $F_oF_1$-ATP synthase in real time during both ATP-driven hydrolysis as well as proton-driven ATP synthesis [4, 13-18]. Attaching one fluorophore on a rotating subunit and a second different fluorophore on a non-rotating static subunit allowed us to measure the distance between these two markers by Förster resonance energy transfer within the enzyme (smFRET). The FRET-labeled single $F_oF_1$-ATP synthase was reconstituted in a liposome, i.e., a 100-nm lipid vesicle. A proton motive force could be generated by a difference in proton concentrations inside and outside of the liposome plus a directional electric membrane potential comprising a $[K^+]$ diffusion potential [19]. Thus, we analyzed rotary step sizes, direction of rotation during ATP synthesis, internal twisting of subunits [20], ATP concentration dependence [17], and the mechanism of ADP inhibition [18].

In the smFRET approach, the liposome with one enzyme diffused freely through the confocal detection volume, limiting the maximum observation times to a few hundreds of milliseconds, and, accordingly, to a few rotary steps per enzyme. In addition, free diffusion caused large fluorescence fluctuations with the photon burst for FRET. As our primary goal is to record sequences of the rotary conformational changes in a single $F_oF_1$-ATP synthase from *Escherichia coli*, we started a collaboration with A. E. Cohen and W. E. Moerner (Stanford University) who presented a technical solution of holding a single molecule or nanoparticle in solution for extended periods of time and with reliable constant photon count rates. Their anti-Brownian electrokinetic trap [21-29] (ABEL trap) uses the position probability information of a fluorescent nanoparticle in a µm-sized laser focus pattern to push back the particle to the target position within the laser focus pattern in real time. A microfluidic sample chamber restricts diffusion into two dimensions, and four electrodes generate the electric fields for moving the nanoparticle. The latest implementation of this confocal ABEL trap with smFRET yields very narrow FRET efficiency distributions, laser power-dependent single-molecule time traces of tens of seconds, and provides diffusion coefficients and electric charge information as well [30]. We could obtain time traces of the catalytic turnover from FRET-labeled single $F_oF_1$-ATP synthase reconstituted in a liposome for several seconds using ABEL-FRET [17, 18, 31].

However, the extended observation times are still not long enough for some slower conformational processes of $F_oF_1$-ATP synthase. For example, the stop-and-go behavior of the enzyme, i.e., being catalytically active for about a second, then lapsing into a temporary inhibited state for a second or longer, followed by another period of activity, cannot be recorded with the high photon count rates needed for precise intramolecular smFRET distance measurements. Photobleaching of either FRET fluorophore is the cause of the limited observation times using conventional organic dyes.

Therefore, we now explore our previously suggested use of a non-blinking, non-photobleaching, bright color center in a nanoparticle in combination with a second label for distance-sensing [32-34]. Both markers will be attached to a single $F_oF_1$-ATP synthase, i.e., one to the rotating and mechanically accessible ring of ten c-subunits in the $F_o$ domain, and the second to non-rotating subunits. The primary marker is the negatively-charged nitrogen-vacancy ($NV^-$) center in a single nanodiamond with less than 50 nm in diameter. Small fluorescent nanodiamonds with average sizes of 10 nm or 40 nm are commercially available, are dispersible in $H_2O$, and are made with a limited variety of surface modifications for specific chemical attachment to the enzyme [35-38]. The $NV^-$ center emits at wavelengths longer than 640 nm, with constant photon count rates, and without blinking or fast photobleaching [39-41].

The additional photophysical benefits of this color center are its fixed nitrogen-vacancy axis within the diamond lattice and the triplet electronic ground and excited states which are affected and can be manipulated by local magnetic fields. $NV^-$ quantum sensing of magnetic fields is achieved by the fluorescence intensity or by the fluorescence lifetime readout, i.e., as an optical detection of magnetic resonance (ODMR) [41, 42]. The $NV^-$ centers in a diamond crystal have four inequivalent crystallographic directions. In a weak external magnetic field, the amplitude of the field and the angle between the N-V axis and the magnetic field determines the microwave (MW) frequencies of transitions between the $m_s=0$ spin state and the $m_s=\pm1$ spin states. Measurements under two or more magnetic fields that were not parallel have determined the three-stepped slow rotation of a single $NV^-$ attached to the thermophilic $F_1$-ATPase domain by fitting the different ODMR spectra with ±4° [43]. Furthermore, precise information about additional pitch and roll motions of the $NV^-$ orientation was obtained. Fast $NV^-$ rotation with more than 3 kHz was resolved by a repetitive read-out scheme using an

NV⁻ center with a long spin coherence time $T_2 > 300$ μs in a diamond with reduced $^{13}C$ content [44]. The rotation speed of the qubit could be detected by quantum state magnetometry [45, 46] and spin-echo interferometry. For the measurements of freely diffusing $F_oF_1$-ATP synthase reconstituted in a lipid nanodisc and held in the ABEL trap, we intend to generate the local magnetic field at the bound NV⁻ center by using an adjacent paramagnetic nanoparticle [47] or ion like $Gd^{3+}$ in a chelate which can influence the three magnetic spin states of the NV⁻ center [48, 49]. Other options to measure intramolecular distance changes within the $F_oF_1$-ATP synthase will include double electron-electron resonance with NV⁻ as well as NV⁻ relaxometry.

Figure 1 illustrates two of the intended experiments on conformational changes of $F_oF_1$-ATP synthase. Precision measurements will reveal the spatial and temporal resolution limits of the NV⁻ orientation marker attached to the $F_oF_1$-ATP synthase. First, the temporal orientational resolution limit is explored by recording the fast ATP-driven 36°-stepped subunit rotation of the $F_o$ motor of the membrane enzyme (ten green helices in Figure 1 A). Second, the spatial or angular resolution limit is revealed by recording of the slow reversible inactive-active transitions of the stator part of the enzyme, and the tilting and rocking motions of the b-subunit dimer (blue or black helices in Figure 1 B). For specific, oriented and rigid attachment of the enzyme to the surface (or to a radical or a superparamagnetic particle, represented by grey blocks at the bottom in Figure 1) as well as to the nanodiamond (represented by light blue cuboids here), orthogonal linker chemistries have to be developed and tested. Therefore, covalent linking by cysteine/maleimide will be combined with the strep-tag/strep-tactin linker system and with the biotin/streptavidin linker system.

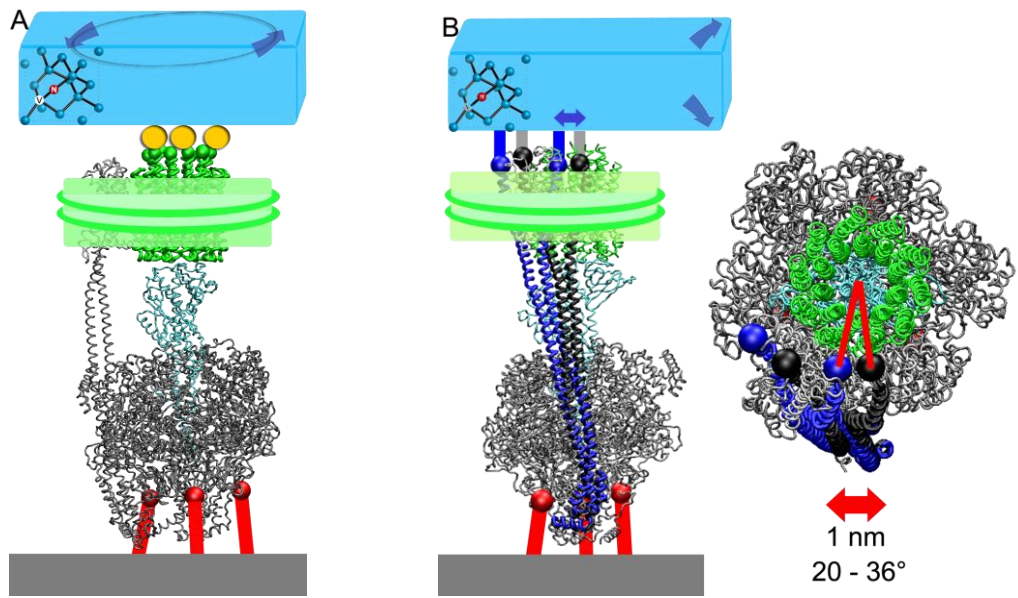

**Figure 1**: Proposed experiments to reveal the spatiotemporal resolution limits of a single NV⁻ in a nanodiamond (blue cuboid on top) as an orientation marker attached to $F_oF_1$-ATP synthase. The non-rotating subunits of the $F_1$ and the $F_o$ motors are shown as grey helices, the rotating γ- and ε-subunits of the $F_1$ motor are shown in cyan, and the rotating ring of ten c-subunits of the $F_o$ motor are shown in green. The enzyme can be attached to a cover glass (grey block below) *via* three linkers (red bars) that are bound to the non-rotation $F_1$ domain. **A**, probing ATP-driven 36°-stepped rotation of the $F_o$ motor (green helices, encircled in a lipid nanodisc symbolized by transparent green block with two peptide rings). **B**, probing the inactive-active transition of the peripheral stator, i.e., the b-subunit dimer (with an enlarged view from above, on the right). The linear distance change of ~1 nm (blue versus black positions of the b-subunit dimer which are shown in two conformations with distinct tilting) corresponds to an angular shift between 20° and 36°.

Nanodiamonds with single NV⁻ centers can be produced by distinct methods [36]. Each batch of an individual nanodiamond preparation will comprise different sizes and shapes which are described by mean values of distributions. The fluorescence brightness of an individual nanodiamond will depend on excitation wavelength and laser power [50], but also on the relative time-averaged rates of the interchanging charge states of the NV center, i.e., either the $NV^0$ state or the NV⁻ state [51]. The surface charge of each nanodiamond will vary, as well as the NV spin states being either $m_s=0$ or

$m_s=\pm 1$. Here we show that we can characterize the heterogeneities of a batch of nanodiamonds, recording one nanodiamond after another and holding it for more than 100 seconds in an aqueous solution by the confocal ABEL trap. Recording individual fluorescence lifetimes with high precision is possible, and a spectral ratio can determined which is correlated to rates of interchanging charge states from $NV^0$ to $NV^-$ and back [52]. From the recorded feedback voltages and the correlated trapping position within the laser focus pattern, we can estimate the diffusion coefficient and, accordingly, the individual size of the trapped nanodiamond. Finally, we apply local magnetic fields and probe the $NV^-$ spin states in solution using fluorescence lifetime measurements as the optical readout.

## 2. EXPERIMENTAL PROCEDURES

### 2.1 40-nm nanodiamonds with one or more NV⁻ centers and coated with streptavidin

We investigated a custom-made preparation of fluorescent 40-nm nanodiamonds with a streptavidin coating (NDNV40nmLwSA2mg, batch ALB-100, Adámas Nanotechnologies, Raleigh, NC, USA) designed for subsequent specific attachment to a biotin-functionalized $F_oF_1$-ATP synthase. The concentration of the fluorescent nanodiamonds was 1 mg/ml in 10 mM Hepes (4-(2-Hydroxyethyl)piperazine-1-ethanesulfonic acid) buffer at pH 7.4. This solution did not contain bovine serum albumin. The number of NV centers per nanodiamond was one to four according to the supplier. A portion of the streptavidin-coated nanodiamonds (7.5 nM) was reacted with 1 μM DOTA-NHS-ester (1,4,7,10-Tetraazacyclododecane-1,4,7,10-tetraacetic acid mono-N-hydroxysuccinimide ester, CheMatech, Dijon, France) in buffer B (10 mM Hepes, 5 mM NaCl, pH 7.5) for 2 h at 20°C, and used without additional purification steps.

### 2.2 Confocal ABEL trap setup with pulsed 531 nm laser

Our FPGA-based confocal ABEL trap setup was described earlier [17, 18, 31, 53-55] and was modified here. Briefly, a fiber-coupled linearly polarized pulsed laser diode emitting at 531 nm with a beam waist of 700 μm (LDH-D-FA-530DL, Picoquant, Berlin, Germany) was used for excitation. The laser repetition rate was set to 10 MHz and the laser power was attenuated to 20 μW before entering the microscope. Laser beam steering was achieved by a pair of electro-optical beam deflectors, EOBDs (Model 310A, Conoptics). A dichroic beam splitter (Z532RDC, AHF, Tübingen, Germany) reflected the laser towards the 100xTIRF oil immersion objective (UapoN100XOTIRF, NA 1.49, Olympus) mounted on an Olympus IX71 inverted microscope. After passing a 200 μm pinhole (Thorlabs), two single photon-counting APDs optimized for lifetime measurements (SPCM-AQRH-14-TR, Excelitas, Canada) recorded the fluorescence photons in two spectral ranges from 540 to 625 nm (BP 582/75, AHF) and for wavelengths $\lambda > 653$ nm (LP 647RU RazorEdge, AHF), separated by a dichroic mirror at 640 nm (ZT640 RDC, AHF). Both APDs and the pinhole were mounted on a single 3D-adjustable mechanical stage (OWIS, Germany). Exact synchronization of the APD signals was achieved by two electronic picosecond delay units ($PSD-065-A-MOD, Micro Photon Devices, Bolzano, Italy). Photons were recorded in parallel on two separate computers. One computer contained the FPGA card (PCIe-7852R, National Instruments) used for the feedback of the ABEL trap, and the second computer was connected to a TCSPC card (SPC-180NX, with router HRT-82, Becker&Hickl, Berlin, Germany). The FPGA Labview program from A. Fields and A. E. Cohen [25] was adapted for trapping on the signals of either one or of both APDs. Adjusting the laser focus position was achieved by a 3D piezo scanner (P-527.3CD with digital controller E-725.3CD, Physik Instrumente, Karlsruhe, Germany). The microfluidic PDMS chip design we used was published previously [31, 56]. To fabricate the disposable PDMS/glass chips for the ABEL trap, the Sylgard 184 elastomer kit (Dow Corning, Farnell, Germany) was used. Short plasma treatment of both the PDMS chamber and the cover glass ensured irreversible bonding.

### 2.3 Analysis of nanodiamond diffusion in the ABEL trap

We recorded the FPGA-generated feedback voltages to the four electrodes of the ABEL trap combined with the FPGA-counted photons from the two APDs. The diffusion coefficient and electromobility were calculated from the reconstructed particle's trajectory based on the photon counts and applied voltages using an assumed density maximum-likelihood estimator [25]. An evolutionary optimization algorithm paired with a simplex-based optimizer improved the identification of the parameters defining the position of the particle in the trapping region of the ABEL trap [57]. Computation over long burst traces took a considerable amount of time and resources; therefore, the particle's trajectory reconstruction was averaged from three to five equidistant 300 ms intervals within a photon burst, depending on its length. Furthermore, each interval was averaged over five fitting iterations to evaluate the convergence of the identified parameters. Outliers, defined as values close to any of the parameter boundaries, were not considered during the averaging step.

The hydrodynamic radius $R_H$ was calculated from equation (1):

$$D = \frac{0.203 k_B T}{\sqrt{6}\, \eta R_H} \tag{1}$$

where $k_B$ is the Boltzmann's constant, $T$ and $\eta$ are the absolute temperature and temperature-dependent viscosity of the medium, respectively.

### 2.4 Analysis of NV$^-$ lifetimes in single nanodiamonds

The fluorescence lifetime of single nanodiamonds was obtained from photon bursts with a minimum length of 100 ms and a minimum number of 7000 photons. Photons were recorded by the TCSPC card. Following photon burst selection in our modified version of the Burst_Analyzer software (Becker&Hickl, Germany), the photon bursts were exported as photon traces to be further processed in MATLAB (R2015a, MathWorks, Inc.). Fluorescence decay histograms consisted of 4096 bins with a bin width of 24.4 ps as defined by the 100-ns micro time of the TCSPC card. Histograms were reconstructed for each photon burst. A fitting region was defined starting 400 ps after the peak of the decay to avoid the contribution of the instrument response function (IRF) and ending at 80 ns. This region was fitted to a triexponential decay function given by equation (2):

$$I(t) = A_0 + \sum_{n=1}^{3} a_n e^{-t/\tau_n} \tag{2}$$

where $\alpha_n$ are the amplitudes of each lifetime component $\tau_n$. $\tau_3$, corresponding to a very short component likely caused by back-scattered laser light (Rayleigh and Raman) and the IRF of laser and APDs, was fixed to 900 ps.

## 3. RESULTS

### 3.1 Brightness and spectral ratio R of individual 40-nm nanodiamonds recorded in the ABEL trap

The nitrogen-vacancy center in fluorescent nanodiamonds (FND) can exist in two charged states, i.e., as the neutral NV$^0$ center or as the negatively-charged NV$^-$ center. Both states are excited into their phonon side bands by the employed pulsed laser system emitting at 531 nm. The two states are distinguished spectrally, with NV$^0$ having a zero-phonon line (ZPL) around 575 nm and an NV$^-$ ZPL around 637 nm [58]. The fluorescence spectra comprise broadened phonon side bands from the ZPL to longer wavelengths causing a significant spectral overlap. Transitions between these states are laser power-dependent for NV$^-$ to NV$^0$ and are reversed spontaneously in the dark within µs [51]. As the ratio between NV$^0$ and NV$^-$ populations also depends on the electrochemical potential at the surface of the nanodiamonds [52], i.e., the surface modification and buffer conditions, we recorded the fluorescence of our streptavidin-coated nanodiamonds with unknown linker chemistry in two spectral ranges. We calculated a spectral ratio R (as described in [52]) from the fluorescence detected between 650 nm and ~800 nm (red intensity trace in Figure 2 A) divided by the summed intensities of this channel plus the intensities from the 545 nm to 625 nm detection channel (green intensity trace in Figure 2 A). The summed photon counts are depicted as the grey time trace in Figure 2 A:

We diluted the FND stock solution (NDNV40nmLwSA2mg, Adámas Nanotechnologies) 50-fold with 10 mM Hepes, 5 mM NaCl, pH 7.5, immediately before transferring the sample into the PDMS/glass microfluidic chamber of the ABEL trap. Thereby, the FND concentration was low enough to have significantly less than one fluorescent particle at a time in the detection volume. Initial ABEL trap parameters for catching the FNDs were set to the 32-point knight's tour laser focus pattern in a 2.34 µm times 2.34 µm trap region repeated with 10 kHz, a diffusion constant D = 15 µm$^2$·s$^{-1}$, a negative electromobility parameter µ = −200 µm·s$^{-1}$·V$^{-1}$ and a beam waist of 0.6 µm. Feedback for trapping was calculated for the combined signals from both APD detectors. With these ABEL trap settings, single FNDs were caught immediately, i.e., one after another and held for tens to hundreds of seconds (Figure 2 A). Almost constant photon count rates for each FND indicated the absence of blinking in the ms range.

In the 65 s section of a 25-min recording, nine individual FNDs were trapped (Figure 2 A). All of them showed different fluorescence intensities and trapping times. FND #1 with a duration of ~14 s had a mean brightness of 93 kHz and a spectral ratio R = 0.89 (with a short reversible change). This nanodiamond was removed actively from the trap by switching of the laser pattern and the electrode feedback voltages, causing a fast-rising intensity peak at the end of the photon burst.

The photon burst of FND #2 with a duration of 2.6 s had a mean brightness of 20 kHz and a spectral ratio R = 0.83; FND #3 a 4.7 s duration, brightness 54 kHz and R = 0.82; FND #4 a 12 s duration, brightness 116 kHz, R = 0.82, active removal from the trap; FND #5 a short 0.6 s duration, brightness 26 kHz, R = 0.80; FND #6 a 1.6 s duration, brightness 56 kHz, R = 0.86; FND #7 a 2.2 s duration, brightness 36 kHz, R = 0.81; FND #8 a 2.2 s duration, brightness 55 kHz, R = 0.78; and FND #9 a 12.8 s duration, brightness 68 kHz, R = 0.86. Figure 2 B summarizes the very different brightness levels and the related spectral ratio values. A straightforward grouping of FND intensities related to either a single NV center or two, three, or more within a single nanodiamond was not possible as a continuum of brightness values exists. However, most FND exhibited intensities between 20 to 60 kHz. Given the laser excitation power of 20 µW at 10 MHz repetition rate, the lowest mean 20 to 30 kHz count rate likely represents the brightness of a single NV center [59].

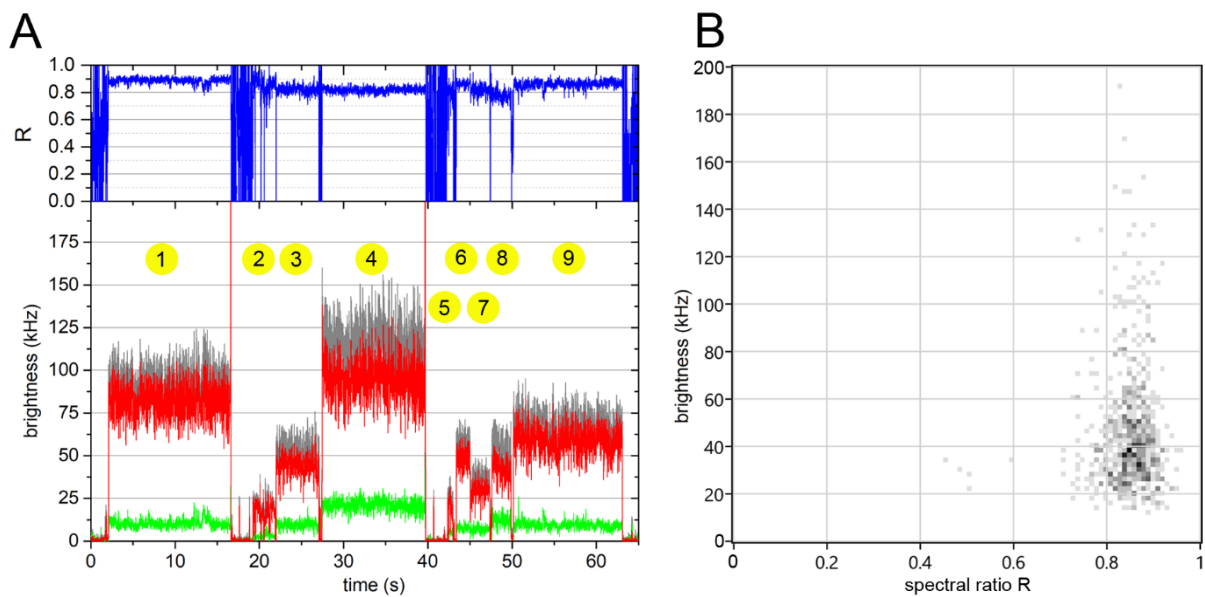

**Figure 2**: Photon bursts from 40-nm FNDs coated with streptavidin and held in solution by the ABEL trap. (**A**) Nine individual FNDs with distinct brightness levels and spectral ratios $R=I_{red}/(I_{red}+I_{green})$ are found in a 65-second section of the recording. Intensity traces are background-subtracted. Spectral detection in the range of 545 nm to 620 nm is shown as the green trace and for λ>650 nm as the red trace. The grey trace is the summed intensity, and the blue trace above is the R time trace. Pulsed excitation with 531 nm at 10 Mhz and 20 µW was used. A 10-ms binning was applied to identify the photon bursts. (**B**) 2D distribution of mean spectral ratio R versus mean brightness of the FNDs. The mean photon counts per 1 ms are plotted as a count rate in kHz.

The spectral ratios from individual FNDs did not deviate significantly from the mean at R = 0.86. The distribution of R values was in very good agreement with similarly produced 40-nm FNDs with carboxy groups on the surface, recorded as immobilized FNDs on a cover glass [52]. Accordingly, our batch of streptavidin-coated FNDs in buffer predominantly comprises NV$^-$ centers, and only five FNDs were found with R values around 0.5 which indicated an NV$^0$ center (Figure 2 B).

### 3.2 Size and charge of individual 40-nm nanodiamonds recorded in the ABEL trap

According to the manufacturer, FNDs were produced by irradiating high-pressure high-temperature microdiamonds with electrons, annealing and milling the obtained microdiamonds, followed by oxidative surface treatment. The very broad size distribution was separated further by centrifugation steps. This yields FND batches with different shapes and sizes as imaged by transmission electron microscopy and characterized by dynamic light scattering [60].

We were interested in determining the size and surface charge heterogeneities of the 40-nm FND with streptavidin coating one by one in the ABEL trap. At first, we recorded intensity time traces of freely diffusing FNDs in the confocal ABEL trap in the absence of a laser pattern and electric feedback. Fluorescence correlation spectroscopy (FCS) revealed a mean diffusion time for the FND which was about 60 to 100 times longer than for the free fluorophore Atto532-NH$_2$ in H$_2$O (Figure 3 A). However, a comparison of several subsequent 60-s sections of the FCS recording revealed large heterogeneities of the respective diffusion times, caused by a few larger and much brighter FNDs in some of the time trace sections.

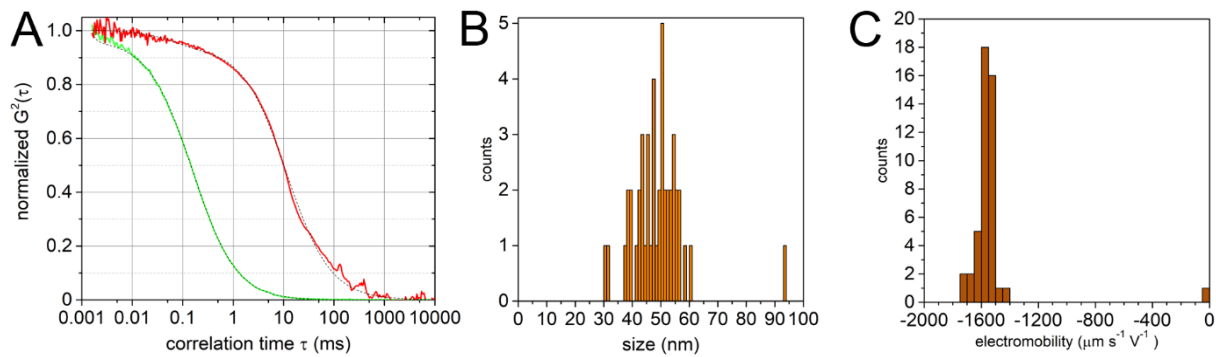

**Figure 3**: (**A**) FCS of freely diffusing FNDs (red curve) and of Atto532-NH$_2$ (green curve) in H$_2$O with the laser pattern and electric feedback switched off in the confocal ABEL trap. (**B**) Size and (**C**) charge distributions of individual streptavidin-coated 40nm FNDs as determined in the ABEL trap.

For ABEL trapping the FNDs, we applied the 32-point Knight's tour laser focus pattern and the electrode feedback to the same solution in the PDMS/glass sample chamber. Electrode feedback was set to ± 10V for x- and y-directions, and the FPGA provided ± 10 V as pulses for each detected photon. The residence time of a laser focus position in the pattern and, accordingly, the fastest electrode feedback time was 3.125 µs. We recorded time traces of the voltage pulses and photon counts in combination with the related Kalman-filtered position estimates and analyzed the diffusion coefficients for each FND as described [25, 57]. A temperature increase to 27° C and the related viscosity decrease of the buffer was taken into account using equation (2) above. From the diffusion coefficient obtained for 300-ms time intervals of the photon burst we calculated the apparent size of the single FND as a spherical nanoparticle. The size distribution is shown in Figure 3 B indicating different diameters ranging from 40 nm to 60 nm for the streptavidin-coated FNDs (nominal mean size of 40 nm). The surface charges were highly negative for these FNDs as represented by the electromobility distribution shown in Figure 3 C.

**3.3 Fluorescence lifetimes of individual 40-nm nanodiamonds recorded in the ABEL trap**

Next, we analyzed the fluorescence lifetime decay histograms for the individual trapped FNDs using a triexponential decay function. Only photons detected in the spectral range from 650 nm to ~800 nm were included. Briefly, we determined the peak maximum and started fitting after a 400 ps delay without deconvolution of the instrument response function (IRF) of the laser pulse and the detection electronics. To achieve precise lifetime analysis, a lower limit of 7000 photons per burst was implemented. One short lifetime component with 900 ps was fixed (but its amplitude was fitted). Two additional longer lifetimes with varying amplitudes were found for each FND. Four examples are given in Figure 4.

The FND shown in Figure 4 A with a size of 46.2 nm exhibited one lifetime $\tau_1 = 4.2$ ns with amplitude $\alpha_1 = 0.31$ and a long lifetime $\tau_2 = 28.4$ ns with amplitude $\alpha_2 = 0.51$. Accordingly, the third fixed 900 ps lifetime component had an amplitude of 0.18. This nanodiamond was trapped for 9.37 s. The total number of photons including background in the lifetime histogram was 277,083 photons, shown as the red histogram in the lower right panel (with fit as a black curve). The mean photon count rate in the detection channel was 44 kHz with background subtracted (red time trace). The FND shown in Figure 4 B with a similar size of 45.3 nm exhibited one lifetime $\tau_1 = 4.9$ ns with amplitude $\alpha_1 = 0.33$ and a long lifetime $\tau_2 = 25.5$ ns with amplitude $\alpha_2 = 0.36$. Accordingly, the third 900 ps lifetime component had a larger amplitude

(0.31). Trapping this FND for almost 23 s yielded 1.66 million photons in the lifetime histogram and a mean brightness of 73 kHz (red time trace).

The FND shown in Figure 4 C with a larger size of 58.4 nm exhibited a short lifetime $\tau_1 = 3.1$ ns with amplitude $\alpha_1 = 0.5$ and a long lifetime $\tau_2 = 18.0$ ns with amplitude $\alpha_2 = 0.41$. Both lifetimes were significantly shortened compared to those of the other FNDs, and the additional contribution of the 900 ps lifetime component was small. The mean brightness of this nanodiamond was low, the trapping time was short (2.1 s), resulting in 42,269 photons in total in the histogram. The mean brightness was 26 kHz. The subsequently trapped FND shown in Figure 4 D was smaller (47.3 nm). It exhibited a short lifetime $\tau_1 = 4.3$ ns with an amplitude $\alpha_1 = 0.41$ and a long lifetime $\tau_2 = 26.3$ ns with an amplitude $\alpha_2 = 0.41$, and a third amplitude of 0.36 for the 900 ps lifetime component. The mean brightness of this nanodiamond was 42 kHz (red time trace) and its trapping time of 1.8 s was short, resulting in 80,402 photons for the lifetime histogram.

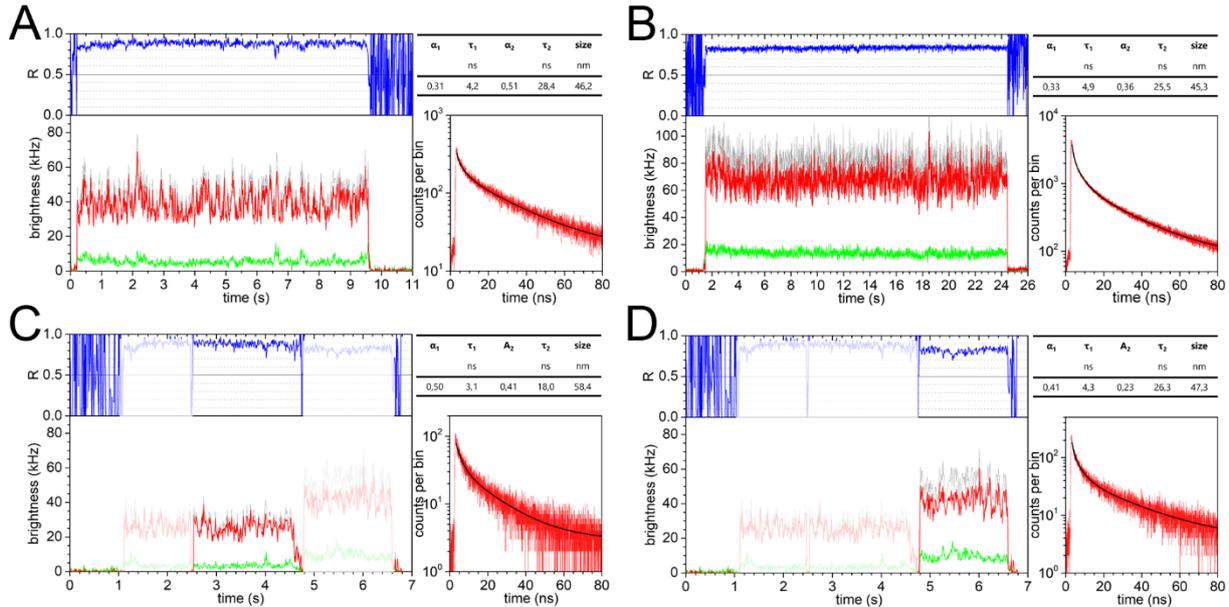

**Figure 4**: Four streptavidin-coated 40-nm FNDs with different lifetimes and sizes, hold in the ABEL trap. Fitting was achieved using a triexponetial decay function. Note that only the two longer lifetimes were compared in the table above the lifetime histogram. A third, very short lifetime with 900 ps was fixed during fitting.

### 3.4 Sensing magnetic fields by single NV⁻ lifetime in the ABEL trap

Next, we modified the streptavidin-coated FNDs (solution with 7.5 nM nanodiamonds) with DOTA-NHS-ester (1,4,7,10-tetraazacyclododecane-1,4,7,10-tetraacetic acid mono-N-hydroxysuccinimide ester, 1 µM), as our goal was to provide several chelate binding sites for $Gd^{3+}$ for future quantum sensing studies. The chemical reaction in 10 mM HEPES buffer was done at room temperature. Unreacted DOTA-NHS ester was not removed afterward. The FND solution was diluted 50-fold in 2 mM HEPES plus 500 µM NaCl at pH 7.5 immediately before transfer into the sample chamber. ABEL trapping was performed with the same initial settings as described above using ps-pulsed 531 nm excitation at 10 Mhz repetition rate.

To evaluate the sensing of a magnetic field using the brightness-independent lifetimes of the NV⁻ center in the FND, we placed a strong neodymium disc magnet in the vicinity of the trap region of the ABEL trap at a distance of ~8 mm. The magnet (NdFeB, N45 magnetization, 35 mm diameter, 20 mm height, weight 172 g; https:\\www.supermagnete.de, Germany) was placed with the round surface to the bottom and the edge facing the PDMS/glass interface of the sample chamber. The strength of the static magnetic field at the trap region remained unknown, but we assumed a strength of about several hundred G. We note that the FNDs are rotating quickly in solution while being held in place by the ABEL trap.

We recorded the FND data sets in the absence of the magnet and then placed the magnet for the subsequent data recording using the same solution and sample chamber. Figure 5 summarizes the distributions of the two fluorescence lifetimes and the total number of photons in the photon bursts in the absence of the magnet (Figures 5 A-C) and in the presence of the magnet (Figures 5 D-F). A significant shortening of longest lifetimes $\tau_{long}$ (relates to $\tau_2$ in the case of streptavidin-coated FNDs without DOTA-NHS, see Figure 4) was observed in the presence of the magnetic field by comparing the yellow bars in Figure 5 A with the red bars in Figure 5 D, respectively. In contrast, the distribution of the short lifetimes $\tau_{short}$ (relates to $\tau_1$ above) was not affected by the magnetic field (grey bars in both histograms).

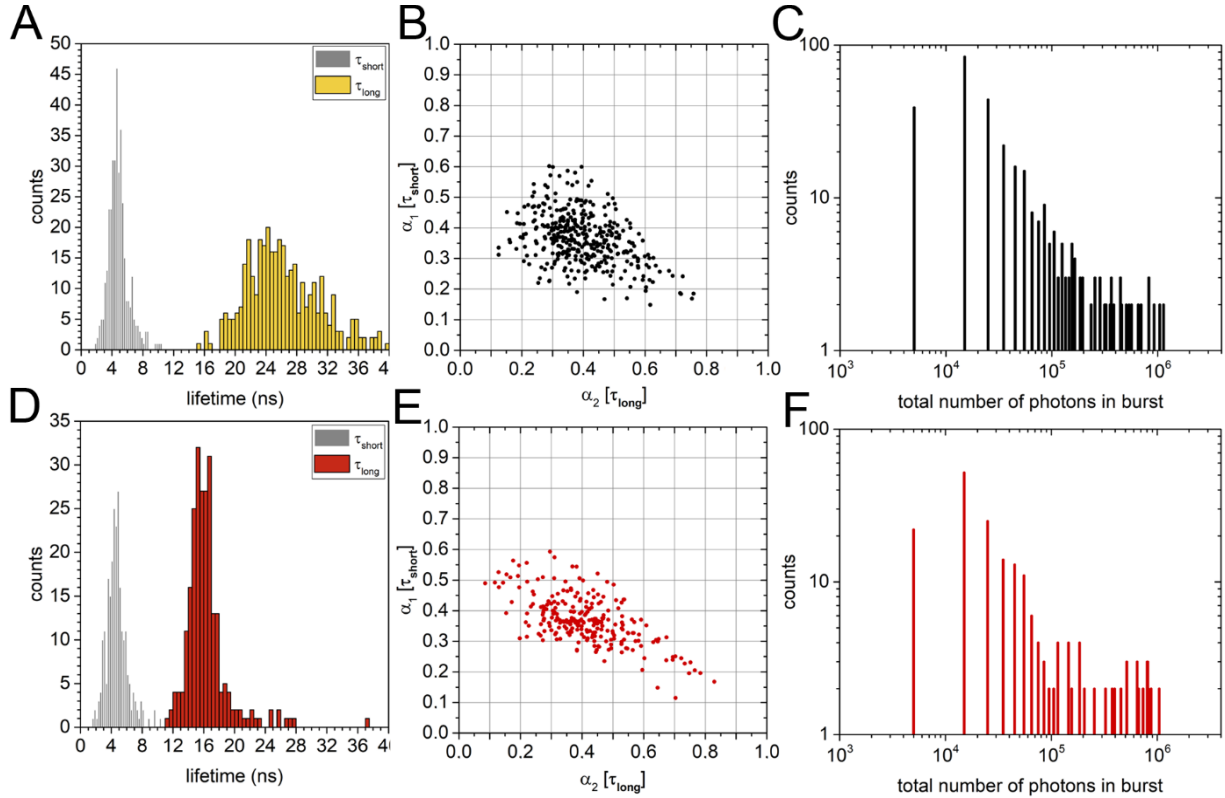

**Figure 5**: Influence of a strong magnetic field on NV$^-$ lifetimes of streptavidin-coated 40-nm FNDs (covalently modified with DOTA-NHS) in the ABEL trap. A macroscopic magnet was positioned at an 8 mm distance from the ABEL trap region. (**A**) Distribution of the two lifetimes in the absence of the external magnet. (**B**) Amplitudes of the ~4 ns lifetime $\tau_{short}$ and the long lifetime $\tau_{long}$ in the absence of the external magnet. (**C**) Distribution of total photon numbers in a burst from an FND in the absence of the external magnet. (**D**) Distribution of the two lifetimes in the presence of the external magnet. (**E**) Amplitudes of the ~4 ns lifetime $\tau_{short}$ and the long lifetime $\tau_{long}$ in the presence of the external magnet. (**F**) Distribution of total photon numbers in a burst from an FND in the presence of the external magnet.

In the presence and the absence of the magnet, the relative amplitudes $\alpha_{1,2}$ of the short and the long lifetimes vary over a very wide range (Figures 5 B, E). All data points near the diagonal from the upper left corner ($\alpha_1 = 1$ and $\alpha_2 = 0$) to the lower right corner indicate a negligible amplitude for the 900 ps lifetime component, whereas those data points in the lower left area, that were located near the diagonal between $\alpha_1 = 0.5$ and $\alpha_2 = 0$ and $\alpha_1 = 0$ and $\alpha_2 = 0.5$, indicated fraction of 50 % for this very short lifetime.

As seen in Figures 5 C, F, the total number of photons from an FND varied over a large range as well, i.e., from 7000 photons set as the minimum to more than 1 million photons in a burst. In the presence of the magnetic field, a trend towards lower count rates was observed. i.e., the shortening of lifetime $\tau_{long}$ was correlated by a reduced brightness as expected for a switch of the NV$^-$ spin state from $m_s=0$ with higher brightness and longer fluorescence lifetimes to the spin state $m_s=-1$ with lower brightness and shortened lifetimes [42].

## 4. DISCUSSION

Fluorescent nanodiamonds with one or more NV centers can be surface-modified for covalent or otherwise strong and specific binding to biological targets *in vitro* and *in vivo*. The fluorescence signal is stable even at higher excitation power, it is almost non-blinking, and also photobleaching is a negligible limitation for very long single FND observation times. However, the production methods for FNDs yield distinct batches of diamond nanoparticles [61]. Individual FNDs differ in size and shape, surface chemistry or the number of functional groups, in brightness, in the atomic nature of the color center, more specifically in the spectral ratio R defined by fast reversible transitions between neutral $NV^0$ and negatively-charged $NV^-$ centers, and in fluorescence lifetimes. NV centers differ in individual spin state properties like the $T_1$ relaxation times, the $T_2$ spin echo dephasing times, possible Zeeman- and Stark-splittings, and other quantum properties due to coupling to neighboring nuclei in the diamond lattice, because the electronic ground states and the fluorescent excited states of the $NV^-$ center are triplet states.

Here we analyzed 40-nm FNDs with streptavidin coating dissolved in diluted HEPES buffer at a physiological pH. Our goal was to map heterogeneities of the photophysical properties by recording time traces of individual FNDs one by one in buffer solution. Extending the observation times to tens and hundreds of seconds of diffusing FNDs in solution was achieved using a confocal, fast anti-Brownian electrokinetic trap as developed by A. E Cohen, Q. Wang and W. E. Moerner. Our version of an ABEL trap is based on the hardware described and their control software provided in 2011 [25], with minimal additions. Subsequent data analysis of the trapped FNDs was also based on their software, with minor additions made to stably fit the diffusion coefficient and electromobility parameter. We confirmed that each FND exhibited a different mean brightness that was not related to its size or surface charges. The mean brightness was stable for most FNDs for many seconds, but we noticed reversible intensity fluctuations (on the hundred ms time scales) up to 30 or 40% in some trapped FNDs. The $NV^0/NV^-$ properties as described by the spectral ratio R differed as well from FND to FND. However, only very few FNDs were unequivocally assigned as nanodiamonds comprising a single $NV^0$ center. We conclude that for the given batch of streptavidin-coated FNDs, the photon recording in a spectral range of $\lambda>650$ nm to 800 nm is sufficient and will enable antibunching or coincidence measurements to determine the number of $NV^-$ centers in each FND in the future.

The two-channel TCSCP recording of photons allowed us to analyze fluorescence lifetimes in combination with the ps-pulsed laser excitation. The trapped FNDs provided tens of thousands and up to millions of photons for the lifetime histogram of a single particle. Therefore, triexponential decay fittings were reliable and necessary. The amplitudes of the three distinct lifetimes varied over large ranges. The significant two (fitted) fluorescence lifetime components were found in the range of about 4 ns and of 20 ns to 30 ns. These two lifetimes were in accordance with published data for related nanodiamonds [62-64]. The very fast fixed 900 ps lifetime component was attributed to Rayleigh scattering, and a possible Raman signal around 1332 $cm^{-1}$ from the diamond lattice [65]. The ~4 ns lifetime was explained by $NV^-$ quenching due to surface properties [66], and the long lifetime to unquenched $NV^-$ away from the surface. As the location of the $NV^-$ center is unknown within the nanodiamond with sizes ranging from 30 nm to 60 nm, the observed large variability of the lifetime amplitudes is not surprising.

How to make use of the photophysical heterogeneities of the FNDs? Importantly, the long lifetime component as a result of the $m_s=0$ spin state can be used to sense a local magnet field. As we have shown here, a strong magnetic field results in a significant shortening of the long lifetime component of $NV^-$. A local magnetic field with appropriate direction with respect to the N-V axis and with optimal strength might facilitate the crossover into the $m_s=-1$ spin state in a distance-dependent manner. The optical fluorescence lifetime-based readout is convenient to implement on a confocal microscope like our ABEL trap. As the fluorescence lifetime measurement is hardly affected by the different brightness of each FND, and the amplitudes of the long lifetime component are in the range of 30 to 60 %, sufficient photons will remain for a precise determination of this long lifetime and any changes of this lifetime.

The long trapping times of soluble single FNDs can be improved further and beyond hundreds of seconds. A higher fluorescence signal using higher excitation power can be achieved and combined with a lowered background using all-quartz sample chambers for the ABEL trap. This will allow us to apply other quantum sensing approaches of local magnetic fields like $T_1$ relaxometry or $T_2$ dephasing measurements, even for FNDs being trapped in solution [67]. As a result of the photophysical heterogeneities, any measurement of dynamic biological systems using a single FND as the reporter has to be complemented by simultaneous recording of the controls (or manipulations) of the specific spin state of the same FND. This is necessary for both types of intended measurements of the conformational dynamics of the membrane enzyme $F_oF_1$-ATP synthase on surfaces and in solution. For the surface-attached enzyme at work, the orientation of the N-V axis in the FND can be followed by three-dimensional magnetometry with very high angular resolution. For the diffusing enzyme in solution with an attached FND and a magnetic nanoparticle as the second marker, we expect that the ABEL trap will guarantee sufficiently long observation times, i.e., likely more than 100-fold longer compared to our previous single-molecule enzymology approaches in the past.

**Acknowledgments**

The authors thank all members of our research groups who participated in various aspects of this work. Funding for this joint research was provided through the innovation project "NV$^-$ qubit for precise sensing of rotational and translational mobilities of active proteins" of the Carl Zeiss Foundation, Center for Quantum Photonics (QPhoton) Jena-Stuttgart-Ulm to A.K., J.W., F.J., and M.B. The ABEL trap in Jena was realized by additional German Science Foundation funds (DFG grants BO1891/10-2, BO1891/15-1, BO1891/16-1, BO1891/18-2 to M.B.). A.K. acknowledges support from the BMBF via project QSENS. F.J. acknowledges support from the BMBF via projects QSENS, MikQSens and Diaqnos, EUREKA via project quNV2.0, QUANTERA via the project "Microfluidics Quantum Diamond Sensor", Carl Zeiss Foundation via IQST and Ultrasens-Vir projects, DFG via EXC 2154: POLiS, SFB 1279 and projects 491245864, 499424854, 387073854, ERC grant HyperQ (no. 856432), EU via H2020 projects FLORIN and QuMicro. J.W. acknowledges the DFG via FOR 2027 and CONFINE, the BMBF via projects QMED, QHMI, NeuroQ, DiaQnos and Quamapolis, and the EU via projects AMADEUS, SPINUS and C-QuENS.